\documentclass[a4paper,floatfix,rmp,twocolumn,showkeys,superscriptaddress,amsmath]{revtex4}
\usepackage[latin1]{inputenc}
\usepackage{graphicx,color,amsfonts}

\begin{document}

\title{Evolving complexity: how tinkering shapes cells, software and ecological networks}

\providecommand{\ICREA}{ ICREA-Complex Systems Lab, Universitat Pompeu Fabra, Dr. Aiguader 88, 08003 Barcelona, Spain}
\providecommand{\IBE}{Institut de Biologia Evolutiva (UPF-CSIC), Pg. Maritim 37, 08003 Barcelona,}
\providecommand{\ETL}{Evolution of Technology Lab, Institut de Biologia Evolutiva (UPF-CSIC), Pg. Maritim 37, 08003 Barcelona,}
\providecommand{\SFI}{Santa Fe Institute, 1399 Hyde Park Road, Santa Fe NM 87501, USA,}
\providecommand{\ECLT}{European Centre for Living Technology, S. Marco 2940, 30124, Venice, Italy}

\author{Ricard Sol\'e\footnote{corresponding author. E-mail: ricard.sole@upf.edu}}
\affiliation{\ICREA}
\affiliation{\IBE}
\affiliation{\SFI}
\affiliation{\ECLT}

\author{Sergi Valverde}
\affiliation{\ETL}
\affiliation{\ECLT}

\begin{abstract}
A common trait of complex systems is that they can be represented by means of 
a network of interacting parts.  It is, in fact, the network organisation (more than the parts) what largely 
conditions most higher-level properties, which are not reducible to the properties of 
the individual parts. Can the topological organisation of these webs provide some insight 
into their evolutionary origins?  Both biological and artificial networks share some common architectural 
traits. They are often heterogeneous and sparse, and most exhibit the small-world property or 
the presence of modular or hierarchical patterns. These properties have often been attributed to the selection of 
functionally meaningful traits. However, the proper formulation of generative network models suggests 
a rather different picture. Against the standard selection-optimisation argument, some 
networks reveal the inevitable generation of complex patterns resulting from reuse and 
can be modelled using duplication-rewiring 
rules. These give rise to the heterogeneous, scale-free and modular architectures observed in 
the real case studies. Tinkering is a universal mechanism that drives not only biological evolution but also the large-scale dynamics 
of some technological designs. Here we examine the evidence for tinkering in cellular, technological and ecological 
webs and its impact in shaping their architecture and deeply affecting their functional properties. 
Our analysis suggests to seriously reconsider the role played by selection forces or design principles as 
main drivers of network evolution.
\end{abstract}



\maketitle


\section{Introduction}

\begin{quote}
\flushright
\em
"Knowing how something originated often\\ is the best clue for how it works" \\
Terrence Deacon
\end{quote}

\vspace{0.5 cm}

The fabric of complexity is made of networks. The presence of collective-level, system properties necessarily 
requires a description grounded in a map of connections between individual parts. Such view has been around much 
longer than is usually acknowledged within the field of Network Science. Long before small worlds and scale-free 
structures were identified, the importance of interactions and their embodiment within graphs was already 
in place in ecology and neuroscience. Classic studies on trophic webs and their stability had an enormous 
impact on our understanding of communities (Margalef 1968, May, 1973, Pimm 1984, Sol\'e and Bascompte 2006). Similarly, 
since Ramon y Cajal (DeFelipe 2006) the realization that cognition was associated to complex webs has been 
percolating through the entire field (Sporns et al 2004, Bassett and Sporns 2017). 

\begin{figure*}[t]
\begin{center}
\includegraphics[width=15 cm]{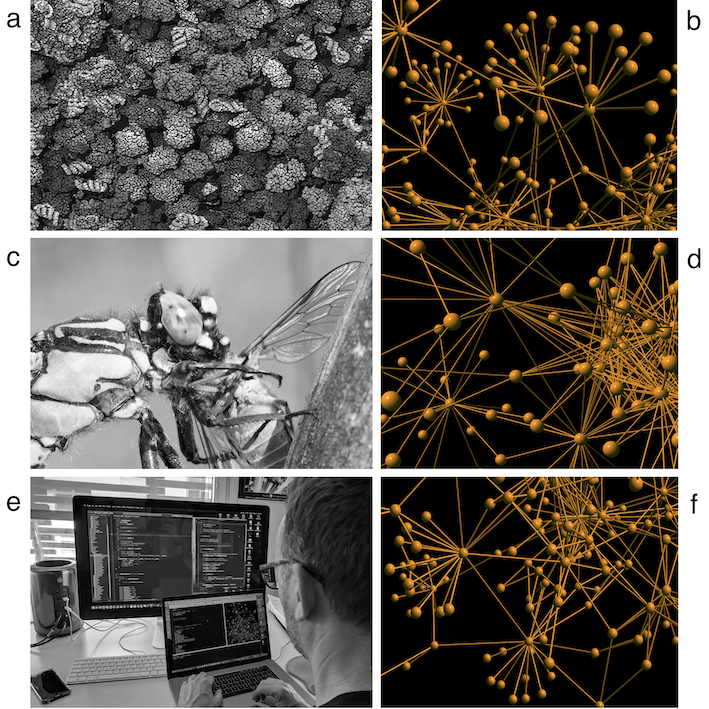}
\caption{{\bf Complex systems and their network architecture}. Three different case studies (left column) include 
(a) molecular interactions within cells (figure adapted from McGuffee and Elcock 2010) which can be described 
for example in terms of protein-protein interaction networks (b) where links indicate physical interactions, Predator-prey 
(c) and other interactions within ecological food webs, define a major class of links within 
complex ecological networks (picture by Dariusz Kowalczyk). An example is shown in (d) where we 
display part of the Silwood Park 
web, where nodes and species and the edges stand for trophic interactions. Finally, artificial 
systems can also be described as networks. A specially relevant example is 
provided by software graphs, created by programmers (e)
who build complex interacting subsystems such as the one displayed in (f). This network is corresponds to the 
Aztec include network (see text) where different parts of the software project and links indicate dependencies between 
files. Although each example result from very different underlying processes and involve 
very diverse scales, they all share a common class of fundamental generative rules associated to reuse.}
\end{center}
\end{figure*} 

 Why to care about interactions among the units? Why is the network level relevant in seeking 
 explanations for the origins of complexity? A main avenue that largely precluded the rise of network science 
 early provided the answers. It was the field 
of phase transitions (Stanley, 1971, Goldenfeld 2018, Sol\'e 2011, Stein and Newman 2013). 
Phase transitions are a common phenomenon that we know well: water freezes or boils 
and completely different, homogeneous macroscopic properties 
can be seen on each phase (liquid, solid and gas) separated by characteristic transition temperatures. But how these 
phase transitions happen? How can we use a microscopic theory that incorporates the properties and interactions 
among components and explain the completely different qualitative properties displayed by each phase? 
 In searching for theories, physicists found that 
very simple models, incorporating a toy description of the interactions (and paying very little attention 
to the structure of the components) can fully explain them. The nature of the interactions, and not the 
components, is what matters. 

The little role played by the details of the units is deeply tied to the presence of emergent patterns arising on the larger 
scale (Anderson, 1972, Kauffman 1993, Gell-Mann 1994) and the failure of the 
reductionist program (Goodwin 1994; Sol\'e and Goodwin, 2001). Emergence became a central concept 
and the cornerstone of a view of systems where system-level properties could not be reduced 
to the properties of its component units. One particularly interesting 
problem that received considerable attention (and that 
will be relevant here) was the phenomenon of {\em percolation} (Stauffer and Aharony 2003): the sudden 
shift from a disconnected to a connected system that is experienced by a set of $N$ elements that are connected with 
pairwise threads with a gi¡ven probability $p$. If $\langle k \rangle$ indicates the average number of connections 
displayed by a randomly chosen element of this system, for a random connection rule we have $\langle k \rangle =pN$ 
and it can be shown that the critical condition $\langle k \rangle_c =1$ defines an abrupt (topological) phase transition (see 
Albert and Barab\'asi 2002). It provides a powerful illustration of emergence and had a tremendous impact 
in very diverse domains, from molecular biology to social dynamics. (Albert and 
Barabasi 2002; Dorogovtsev and Mendes, 2002; Newman 2003; Bocaletti et al 2006; Estrada 2012; Barabasi, 2016). 

With the turn of this century, the combination of available data sets 
associated to social, technological and biological systems and new generation of theoretical models triggered 
a revolution that spread across all disciplines. Network Science became the 
basis for both new concepts and statistical methods. 
Classical graph theory (Gross and Yellen 2005) had been mainly associated to the study of homogeneous webs, including 
in particular all kinds of lattices, trees and random graphs. Because the lack of knowledge about 
the architecture of real networks, most models of genetic, neural and ecological systems 
used homogeneous graphs. Everything changed once the mapping of Internet, social webs or -omic data provided 
the first solid picture of real networks. They confirmed an old sociological study (Milgram 1967) indicating that 
the number of degrees of separation between two randomly chosen individuals, providing a 
very elegant mathematical framework to explain the phenomenon (Watts and Strogatz 1998). Secondly, 
an unexpectedly general finding was the realisation that real networks are heterogeneous. Instead of 
previously assumed regular, Gaussian or exponential distributions of connections, long-tailed 
ones were found to be the rule: most elements are connected with a few others whereas a small number of them 
includes units connected to many others (the hubs).

It has been suggested (Sol\'e et al., 2003) that inspecting the organisation of 
complex networks can reveal the evolutionary design or evolutionary principles 
that shaped them. In a nutshell, identifying the generative rules responsible for their topology 
could be used to find their origins and the contributions of randomness, architectural constraints or 
self-organisation. In other words, the paths followed by each system (either through evolutionary dynamics, 
developmental processes or technological design protocols) can be deeply limited by 
fundamental principles of mathematical nature. Very often, this reveals the failure 
of some standard assumptions and explanations involving selection or optimisation. 

In this paper, we will review the existing evidence for this idea and its deep implications 
for our understanding of network complexity. This includes: (1) the presence of 
mechanisms of network growth that are dominated by extensive reuse of extant components. Such "tinkering" process (Jacob, 1977) was early suggested by the French biologist Francois Jacob and has enormous importance in evolution. Evidence from multiple examples  support this conjecture, but network structure provides a systems-level confirmation of its impact (including engineered, man-made systems). 
(2) Structural constraints limit the repertoire of potential forms of network complexity that can be found. (3) Some particular structures (at different scales) display robust statistical patterns that could be attributed to optimal or efficient design principles: they are byproducts of the heterogeneities created by generative rules.

As will be shown below, deep connections exist between the 
structuralist view of biology, the tinkered nature of complexity and the ways 
networks fill the space of the possible.


\section{Evolutionary tinkering in cellular networks}


\begin{quote}
\flushright
\em
"Living organisms are historical structures (...) 
They represent not a perfect product of engineering, but a patchwork of 
odd sets pieced together when and where opportunities arose." \\
François Jacob
\end{quote}

\vspace{0.5 cm}

 \begin{figure*}[ht]
\begin{center}
\includegraphics[width=16 cm]{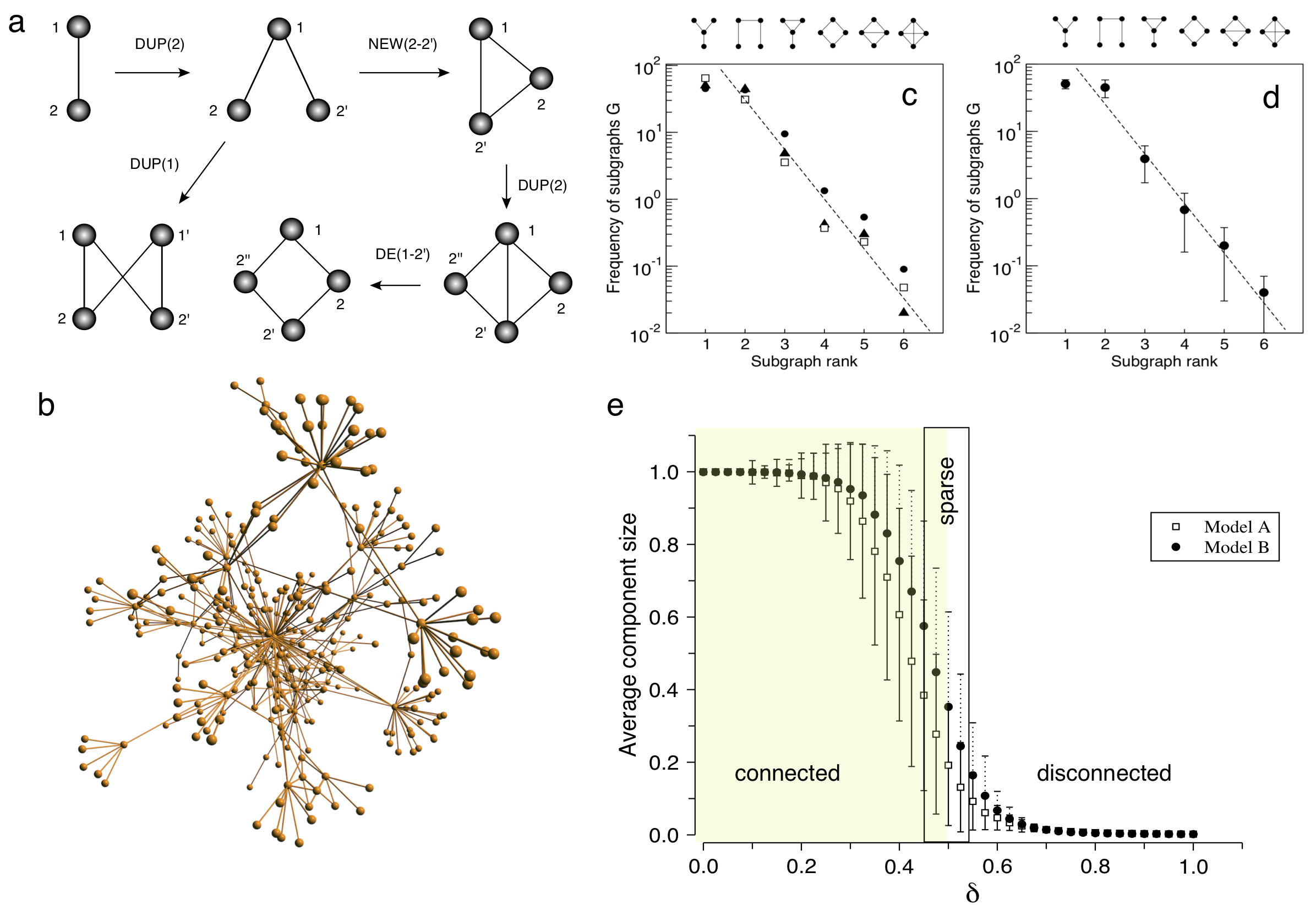}
\caption{Modeling tinkered evolution in complex cellular networks. 
Simple rules of reuse (a) generate heterogeneous networks closely similar to those found in 
cells, such as the Human transcription factor network shown in (b), described in (Rodriguez-Caso et al. 2005). 
Here hubs include key TFs (such as the tumor supressor P53) associated to proliferation diseases. 
Most measurable network properties found in proteomes can be replicated, 
such as the distribution of network motifs. In figure (c) we display the observed frequencies (ordered from 
or three different protein networks, namely the human interactome (HI, filled triangles), 
yeast proteome (YP, filled circles) and the subset of human transcription factors (HTF, open squares). For comparison we display the 
predicted subgraph census from the duplication-rewiring model (d, see Sol\'e and Valverde 2008). By tuning the removal rate of links 
associated to divergence, it can be shown that a transition from connected to disconnected occurs close to 
a critical value $\delta_c=1/2$, as shown for example from the estimation of the average network size against $\delta$ (e). 
Here two different models (Sol\'e et al 2003, Vazquez et al 2003) have been used, here indicated as A and B, respectively.}
\end{center}
\end{figure*}

Our first approach to the evolution of complex networks will consider the impact of a widespread 
feature of evolutionary dynamics: tinkering. As discussed in (Jacob, 1977) 
there's a major difference between the ways an engineer and evolutionary processes work. The first has a 
goal and thus foresees what needs to be designed. To do so, the engineer can completely replace 
previous components and ignore previous protocols to build an artifact. In this way, 
novelties can emerge as a result, for example, of adopting a new material (such as liquid crystal). 
In stark contrast, evolution does not foresee what solution is required for a given problem. Moreover, 
there's no way evolutionary processes ignore those components and structures that are already 
in place: reuse of previous parts is crucial. Surprisingly, reuse leads to network heterogeneities that 
strongly  constrain the local and global properties of the resulting graphs. 

An example of the implications of tinkering rules can be found in models of protein interaction networks. 
As with other non-directed graphs, we define a network $\Omega$ as a pair $(V,E)$. Here $V$ stands now for the 
set of nodes and $E$ the matrix of interactions between pairs of nodes. If we take a given protein $v_j \in V$ 
it will have a given number of links with other proteins in our graph. 
These networks are characterised by the presence of degree distributions following a truncated 
power law shape, i. e. 
\begin{equation}
P(k) \sim (k + k_0)^{- \gamma} e^{k/k_c}
\end{equation}
where $P(k)$ is the probability distribution of having a protein interacting with $k$ other proteins.

Consider the following model (Sol\'e et al 2003, Pastor-Satorras et al 2003). The protein network is now reduced to a graph defined by a collection of 
vertices $V=\{ v_k \}$ and a matrix of connections provided by the adjacency matrix $A=(a_{ij})$ where 
$a_{ij}=1$ if the two proteins interact and zero otherwise. Let us indicate by $\vert V \vert$ the size of the 
network at a given step. We start form a set $m_0$ of connected nodes, and at each time step we
perform the following operations (figure 1a): (i)  One node of the graph is selected at random and duplicated; 
(ii) The links  emanating from  the newly  generated  node are
  removed with probability $\delta$ and (iii) New  links (not previously present  after the duplication
  step) are created  between the new node and all  any other node with
  probability  $\alpha$.  Although available  data  indicate that  new
  interactions are likely to be formed preferentially towards proteins
  with high degree here we do not consider this constraint. 

Step (i) implements  gene duplication (the tinkering component) in which both  the original and
the  replicated proteins  retain the  same set of  interactions. The rewiring  steps 
(ii) and  (iii) implement mutations which translate  into the deletion or addition of  interactions with
different proteins. The  process is repeated  until $N$ proteins have been obtained. 
An example of the networks obtained from this process give networks is shown in (figure 2b) closely 
resembling the protein maps found in cells (fig. 1b). This network fits very well the heterogeneous degree 
distributions, but also other distributions associated to small-scale patterns, such as the abundance of three- and four-component (fig 2 c)
network motifs (Milo et al., 2002) which naturally result from our tinkering 
process (see the subgraphs shown in figure 2a). The model correctly reproduces the observed data (fig. 2d). 
In general, this model and the work by Vespignani and 
co-workers (Vazquez et al, 2003; Colizza et al 2005, Vazquez 2003) as well as duplication models involving gene 
regulation (Teichmann and Babu 2004) successfully 
explain the architecture of protein maps (van Noort et al 2004; Arabidopsis Interactome 
Mapping Consortium, 2011).

The network shown in fig. 2b is sparse, but the model described above can give highly connected 
or even totally disconnected graphs. Deletion rate in particular needs to be tuned. What can the model 
say about this? It is not difficult to show (Sol\'e et al 2001) that the rate of link removal defines two phases 
separated by a critical value $\delta_c$. This is obtained from an average (mean field) description 
that gives a dynamical equation for the average degree:
\begin{equation}
{dK(n) \over dn}= {2\alpha \over n}+  {1-2 \delta \alpha \over n}K(n)
\end{equation}
where $n$ is the proteome size (acting as a time scale). By solving this equation, a steady state $K^*$ 
is obtained, namely: 
\begin{equation}
K^* = \lim_{n \rightarrow \infty}K(n) = {2 \alpha \over 2 \delta -1}
\end{equation}
One phase, associated to deletion rates $\delta>\delta_c$, involves a 
disconnected proteome (fragmented into different parts) and there is another phase where it becomes 
rapidly hyperconnected. At the critical point separating these two phases (figure 2e) we see the emergence of a sparse, but 
connected graph associated to a percolation transition. In other words, there is 
a point where the protein interaction network becomes a true system (with paths connecting 
all elements) at a low cost. This is the region where the model fits very well the data in the two main models 
of duplication and divergence. 

Perhaps more surprising, the networks 
resulting from tinkering also display modularity: they appear organized in subsets whose elements 
are more connected among them than with the rest of the web (Sol\'e and Valverde, 2008). This 
was unexpected, since modularity is considered one of the most important patterns associated to 
adaptive, functional traits of biological complexity (Raff 1996; Hartwell et al. 1999; Wagner et al. 2007). 
Multiple cellular functions are carried by subsets of proteins, including developmental modules. In general, biological modularity provides the right trade-off 
between specialization and integration (Wagner 1995; Calabretta et al. 2000; Lipson et al. 2002). 
Against that intuition, it was shown that modularity can emerge {\em for free}. As it occurs with other network patterns to be discussed below, 
it might be the inevitable outcome of reuse. Similarly, a common statistical pattern is also 
found at the level of so called network motifs (Milo et al 2002) which were. already known from early work on 
social and ecological webs (Stone et al 2019).

The topological structure of protein nets, as discussed above, might be largely the 
by-product of the network construction process. Tinkering is an inevitable constraint in the 
process of increasing complexity through genome expansion. This is of course 
one part of the whole range of possible changes affecting evolving genomes. Other 
processes such as de novo gene formation or whole genome duplications would 
influence the outcome (Roth et al 2007). But the main driver shaping the interaction network 
remains the same (Middendorf et al. 2004).

There is an important consequence of these results within the context of evolutionary theory. 
The observed patterns, as mentioned above, describe a set of correlations that require an explanation. 
The network organization found in these graphs include structural traits that could be interpreted 
as resulting from selective pressures. But sometimes structural patterns have no adaptive 
meaning. They are actually examples of what Stephen Jay Gould and Richard Lewontin called 
{\em spandrels} (Gould and Lewontin 1979). The term spandrel, borrowed from the vocabulary of 
architecture by these authors, defines the space between two arches or between an arch and 
a rectangular enclosure. In evolutionary biology, a spandrel is a phenotypic characteristic that 
evolved as a side effect of a true adaptation (Sol\'e and valverde 2006). The features of evolutionary spandrels can be 
summarised as follows:
(a) They are the by-product of building rules. They have intrinsic, well-defined, non-random features.
(b) Their structure reveals some of the underlying rules of the system's construction.

The findings described above strongly support 
the idea that the proteome map contains a considerable amount of statistical correlations 
that are a byproduct of the duplication-rewiring set of rules.  The presence of modularity or non-random 
distributions of motifs cannot be taken (alone) as a signature of selection. This finding, as shown below, is far from accidental.


%
\section{Software networks: human tinkering}
%


\begin{quote}
\flushright
\em
"A complex piece of software is, indeed, best regarded as a web that has been delicately pieced together from simple materials"  \\
Donald Knuth
\end{quote}

\vspace{0.5 cm}


Technology, as we mentioned above, is different from biology in a fundamental way. 
Jacob's conjecture about this seems very well established. 
There is no intentional design in evolution, 
but it turns out that the limitations imposed by technological complexity 
imposes an effective design barrier to engineers. 
To illustrate this idea, we will use the field of software engineering. 
The reason of this choice is  twofold. One one hand, software projects allow to
extract a network of interactions among different parts in a meaningful way, 
Secondly, software has become increasingly relevant since its inception in the 1950s and has become, even more than the Internet and the Power grid,
 the most important (but largely invisible) technology 
 (Valverde et al. 2002, Valverde and Sol\'e 2003, 2005a, 2005b; Valverde and Solé, 2015; Valverde, 2016). 

Like many technologies, a software system consists of a large number of elements interacting with each other.  Software is similar to a collection of linked pages in a website, but instead of being readable by humans, they are written in a programming language meant for computer processors (Friedman et al., 2008).
Consider the so-called {\em dependency graph} $G = (V, E)$ where a vertex $v_i \in V$ is a program file (e.g., written in a programming language like Java, C or C++). The links between files are described by a (directed) edge $e_{ij} \in E$ indicating a code dependency between source files $v_i$ and $v_j$  which is encoded with the reserved keyword {\em include} (Lakos, 1996). No other information is considered by the network reconstruction algorithm, which means that we are not including any further information, such as functionality. We keep ourselves in the purely topological level. Using this network reconstruction, it was shown that different large-scale software graphs shared a surprisingly universal set of traits.

\begin{figure*}[t]
\begin{center}
\includegraphics[width=0.8\textwidth]{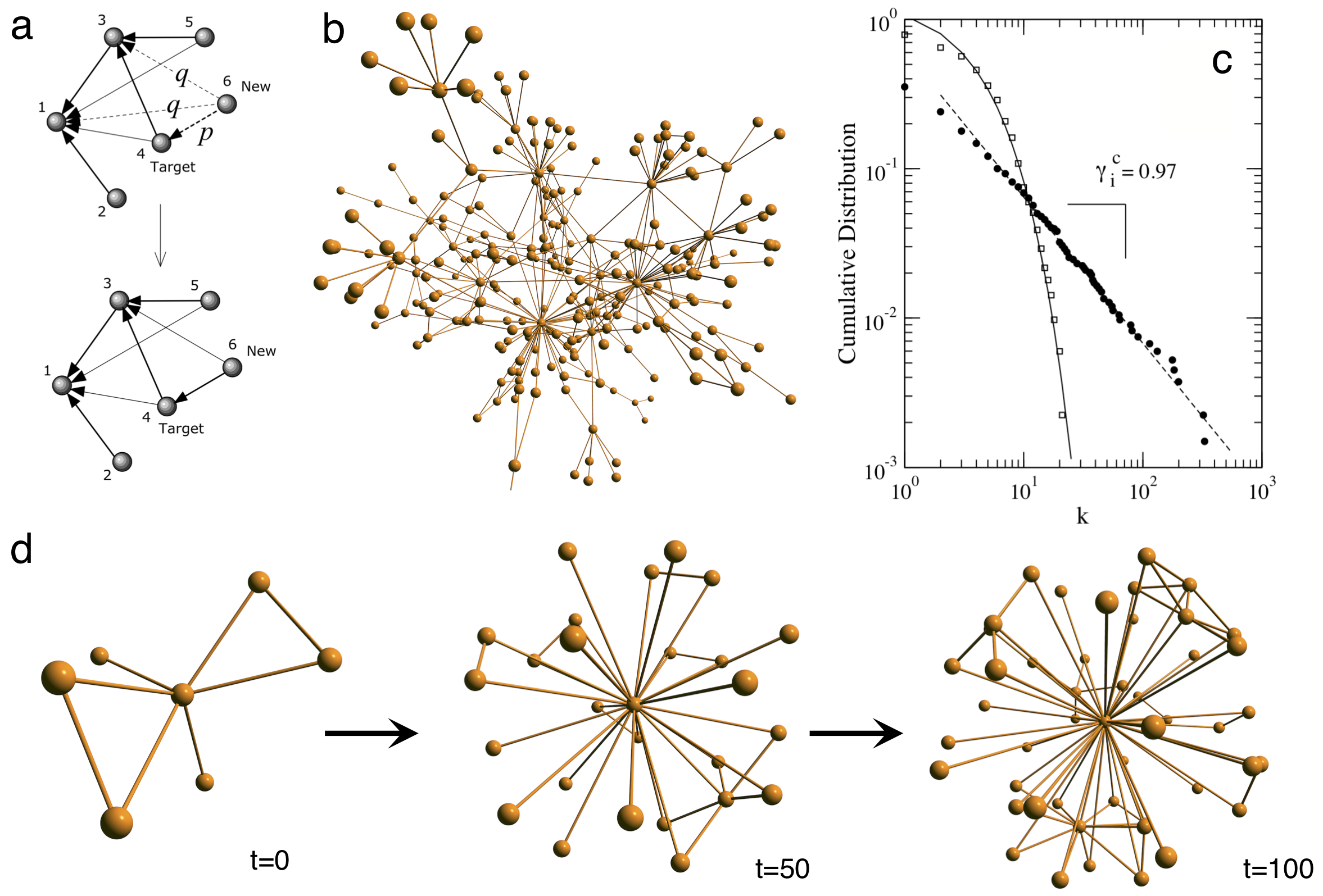}
\caption{ Modelling software dependency graphs.  (a) A simple growth model simulates tinkering in software development processes. A programmer creates a new file by inheriting source code (and their dependencies) from other,  existing $m$ source code files or targets (each target file is referred with probability $p$). Afterwards, code editing removes some dependencies from the new node with probability $q$. Our model generates networks which look like real dependency graphs. For example, (b) the software architecture of XFree86 (a very popular Unix package developed since 1992, displays scale-free behaviour. In (c), the  cumulative distributions associated to $P_i(k)$ and $P_o>(k)$ are shown. The power-law fit of the cumulative in-degree distribution follows a power law with ${\gamma_i}^c \approx 1$ while the out-degree distribution is exponential. (d) From left to right, we show the temporal evolution of a software network generated with tinkering rules. In spite of not assuming any kind of function, duplication and rewiring rules predict the heterogeneity of real dependency graphs, the emergence of clusters of source code files, and even the temporal evolution of the average connectivity.}
\label{fig:aztec}
\end{center}
\end{figure*}

One in particular is the marked difference and statistical nature of the in- and out-degree distributions (Valverde and Sol\'e 2005). These distributions, to be indicated as $P_i(k)$ and $P_o(k)$, respectively, indicate the probability of having $k$ links coming to or leaving the node, respectively. They are very different: the first follows a power law, i. e. $P_i> (k) \sim k_i^{- \gamma_i}$ 
whereas the second follows an exponential decay, as displayed in  figure \ref{fig:aztec}b with the cumulative distributions. In order to compute them, we just need to integrate the degree distribution $P(k)$, i. e. 
\begin{equation}
P_>(k) = \int_k^{K_m} P(k') dk'
\end{equation}
where $K_m$ is the largest degree value. If the original distribution is already a power law, i. e. $P(k)=k^{-\gamma}$ it is easy to show that, for large $K_m$, $P_>(k)  \sim {k^{\gamma-1}/(\gamma-1)}$ The fit to the exponent is particularly good for these fat-tailed distributions using the accumulated form.  For the software graphs discussed here, the in-degree distribution has a characteristic value very close to $\gamma \approx 2$ in all the analysed systems (Valverde and Sol\'e, 2005b).  This is consistent with the pervasiveness of scaling laws in software graphs, which seems largely independent of the choice of programming languages and/or specific functional traits (Valverde et al. 2002; Potanin et al., 2005;  Concas et al., 2007; Louridas et al., 2008).

What is the origin of these patterns? Why are they so close, despite the obvious dissimilarities to be expected from different engineering projects? 
The response seems to be, again, tinkering. Software projects, as much as they are planned and serve to a given objective, are developed by engineers that create a system that grows over time. Diverse parts of the system are created by programmers with a good picture of the global organisation and the parts to be assembled. This assembly is assumed to follow rational design principles (Pressman, 2009). But, as the network grows, that knowledge is eroded and interactions among components are more unpredictable leading to conflicts between subparts. A common strategy is to reuse extant parts of the network and use them as the starting point for a new component.  

Code duplication in software projects is much more common than one would expect. Reuse is a natural process during development of complex software, where new code shares common traits with previously created elements. Duplicate code is reused at different scales, including large blocks containing millions of lines of code. Useful blocks deal with specific tasks and they are often encapsulated in well-defined software modules. Internally, these modules comprise sets of densely interconnected software elements (Subelj and Bajec, 2011; Concas et al., 2017). Since they are sophisticated structures, once a block has been created, a large part of its structure is more likely to being reused in the future. As the system becomes more complex, programmers rely more frequently on duplication (also known as cloning), since they know from experience that these structures are quite stable and  behave well. Empirical analysis of software evolution confirms that duplicate codes are more stable than non-duplicated codes (Krinke, 2008). 

A simple model, known as the Growing Network with Copying (Krapivsky and Redner, 2005) 
explains the asymmetric nature of the two degree distributions and the robust scaling exponents. The rules 
are summarised in fig, 3a and include directed links. The network grows by introducing a single node at a time. This new node links to $m$ randomly selected target
node(s) with probability $p$ as well as to all ancestor nodes of each target, with probability $q$. Specifically, if we label 
each node with a number indicating its age (number one is the oldest). A new node (number 6) attaches to
target node $4$ with probability $p$. This new node inherits every link from the target node (dashed links), with probability $q$. The corresponding mathematical model for the average number of links $L(n)$ is now
\begin{equation}
{dL(n) \over dn}= mp +  mq{L \over n}
\end{equation}
Once solved, it gives us the average degree $\langle K_n \rangle = L(n)/n$. Interestingly, the solution to this equation 
gives three main regimes: (a) a constant average degree $$\langle K_n \rangle \sim {mp \over 1-mq}$$ 
when $mq<1$, (b) a rapidly growing phase with accelerated 
increase $$\langle K_n \rangle \sim N^{mq-1}$$ (ending up in a fully connected web) when $mq>1$ 
and (c) a {\em marginal} state given by $mq=1$ that  
displays a logarithmic growth, i. e. $$\langle K_n \rangle \sim Ln N$$
Interestingly, software graphs seem to grow 
right at this critical point (Valverde and Sol\'e, 2005) separating no increase in the number of dependencies from a super-connected system. 
As shown by Kaprivski and Redner, the model predicts the previously described distributions at the marginal point i. e. 
$P_i(k) \sim k^{-2}$ and an exponential decay for $P_o(k)$. 
 
The universal class of structural patterns obtained by duplication rules is consistent with the hierarchical architecture of software networks. Although human minds take part in the manufacturing of software technology (instead of the blind watchmaker) and software systems cannot repair themselves (Nehaniv et al., 2006), tinkering appears to be the main driver for the large-scale software structure. In addition, software design principles include fault-tolerance and extensibility, and the latter is closely related to the capability of software to evolve. The fact that a simple growth model is able to explain most of the statistical properties of software graphs (Valverde and Sol\'e 2005a, b) is an unexpected turn to Jacob's views: even the engineer, who consciously plans for the future, cannot avoid tinkering when preserving software stability.

We can find common signatures of tinkering in the statistics of software subgraphs. As it happened with network motifs for 
cellular networks, here too, the distribution of motifs in dependency graphs is essentially a byproduct of tinkering.  Given a (sparse) subgraph motif $\Omega$, the average number of occurrences $\left < G \right >$ of  this motif in an uncorrelated scale-free network scales with the subgraph size and the exponent of the in-degree distribution (Itzkovitz et al., 2003): 
\begin{equation}
\left < G \right > \sim N^{n - g + s - \gamma_i + 1}
\end{equation}
where $s$ is the maximum in-degree in the motif and $n$ and $g$ are the number of nodes and links in the motif, respectively. This scaling is actually valid for $2< \gamma_i < s + 1$.  Interestingly, this prediction is in very good agreement with the number of appearances of motifs in software networks (Valverde and Sol\'e, 2005b). 

It has been proposed that network features (e.g., motif abundances) reveal functions (Milo et al., 2002). However, software and proteome networks display very similar structural features, including scaling exponents, motif distribution and modularity, which can be readily explained using duplication-rewiring  mechanisms. The remarkable convergence between natural and artificial networks suggests the origin of motif abundances is neutral to selection. Assigning specific functions to motifs does not take into account the plasticity of complex systems. Indeed, the same network motif can perform many different functions depending on  global requirements (Valverde and Sol\'e,  2013; Widder et al 2012).  In this context, the origin of universal network patterns might be largely independent of specific selection pressures, but this does not mean they are not advantageous.  

Many other relevant findings emerged from the first study of software maps (Valverde and Sol\'e, 2007). One of them was the concept 
of {\em breakdown of modularity} (Valverde, 2017): because of a tendency of software projects to become more disordered (in terms of the pattern 
of dependencies) modules can become more interconnected. Because of this, changes propagate (Valverde, 2007). Within the context of 
software engineering, shortcuts among modules would allow changes made in one module to other, apparently distant parts. These kind 
of phenomena were afterwards described within the context of brain networks, where nowadays the concept of disconnection disease 
(Friston 1995) has
become a central one. Breakdown of modularity 
would be key to understand faulty intermodule communication in e. g. Alzheimer's disease or schizophrenia  
(David, 1994, Alexander-Bloch et al., 2010, Meunier et al., 2010, Bashan et al., 2012; Godwin et al., 2015).


%
\section{Evolution of ecological tinkered networks}
%


\begin{quote}
\flushright
\em
"It is interesting to contemplate a tangled bank (...) and to reflect that these elaborately constructed forms, so different from each other, and dependent upon each other in so complex a manner, have all been produced by laws acting around us."  \\
\vspace{0.15 cm}
Charles Darwin
\end{quote}

\vspace{0.5 cm}

 \begin{figure*}[t]
\begin{center}
\includegraphics[width=0.75 \textwidth]{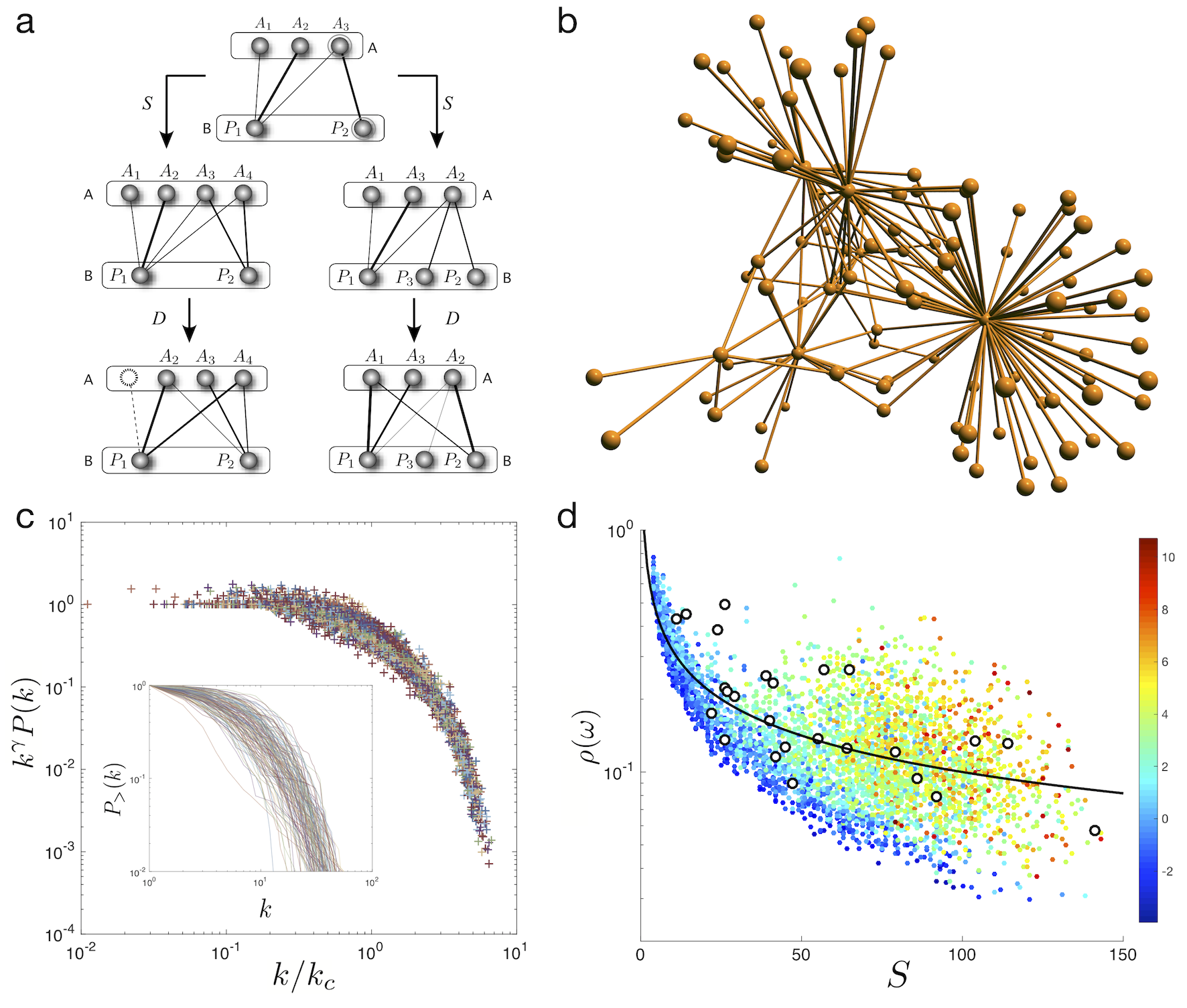}
\caption{ Modelling tinkering in ecological networks. On a large scale, involving whole ecosystems, rules of tinkering can be understood in terms of species-level evolutionary events. (a) In particular, the large-scale 
organisation of mutualistic webs can be obtained from a simple model where speciation (S) of  extant species is the equivalent to duplication events, whereas species divergence (D) is here related to link weight redistribution over evolutionary time. Using this model, the resulting webs (b) display truncated power laws (c) fully consistent (in average and diversity) with observable ones.  (d) The spectral radius $\rho(\omega)$ (a quantitative measure of nestedness) is plotted against the network size ($S$) for a large sample of random networks evolved by duplication and divergence (node colour represents the statistical significance of the measured nestedness, higher values are more significant). Open circles correspond to real networks, including the one displayed in the panel (b), while the black line defines the analytical prediction for the random model. }
\label{fig:mutual}
\end{center}
\end{figure*}

On a much larger scale, consider the large-scale organization of ecological systems, as described in terms of complex networks. There is a long tradition within ecology where networks of some kind have been studied. The reason is obvious when we realize that ecology has always been a systems science, with ecological interactions placed at the center of any meaningful description of communities and their environments (ref). However, the rise of network science provided a novel view of the fabric of nature. The early papers on ecological networks (Sol\'e and Montoya 2001; Montoya and Sol\'e 2001; Montoya et al., 2006, Dunne et al., 2002) soon indicated that they shared the complexity-fragility trade-offs with other complex systems. A very important implication of that was the existence of network co-extinctions and cascades promoted by the propagation of species losses through the trophic structure (Sol\'e and Montoya 2001). Some simple, elegant models revealed, here too, that simple rules could help understanding the origins of food web graphs (Williams and Martinez, 2000; Camacho et al 2002a, 2002b). 

Along with predator-prey interactions, which include carnivorous interactions that benefit one organism (predator) to the detriment of another (prey), another important class of exchanges deals with cooperative interactions. This is the case of {\em mutualistic} webs, which are associated to pairwise relations where the two species help each other (Bascompte and Jordano, 2007; Bastolla et al., 2009). This can occur through a diverse number of ways, including pollination (transfer of pollen from a male plant to a female plant by another agent, including animals, insects and birds) or through frugivory (dispersal of seeds along the fruits consumed by an agent). In both cases, it is assumed that a long-term coevolutionary dynamics has been shaping these interactions (review). Some examples that are often mentioned illustrate this coevolutionary path with specialized pairs. But a more appropriate picture tells us that there is a range of possible patterns of interaction where generalists also abound.

The architecture of mutualistic webs has been shown to display well-defined, common structural regularities already discussed above. 
In contrast with previous examples, a bipartite organisation is present: there are two distinct classes of organisms (such as plants and their pollinators). 
They are small worlds (Ollesen et al., 2006; Montoya and Sol\'e, 2002;  Watts and Strogatz, 1998)  and the distribution of connections is known to follow a truncated power law shape:
\begin{equation}
P(k) \sim k^{-\gamma} \exp (-k / k_c ) 
\end{equation}
where the exponent $\gamma$ for this distribution has a value close to one. Moreover, several additional structural patterns have also been found. These include a highly asymmetric matrix of interactions (Bascompte et al., 2006) and, more importantly, the presence of nestedness (Bascompte et al., 2003). In a nested network, both generalist and specialist are more likely to interact with generalists while interactions among specialists are rare. Nestedness has been shown to positively correlate with biodiversity (Bastolla et al., 2009; Saavedra et al., 2016).

We ask whether speciation and divergence rules alone can explain structural regularities of mutualistic networks. This approach makes strong assumptions. First, we assume that species are present or absent and their abundances (and other species-specific traits) do not play any role. Secondly, weighted links represent quantitative interactions among species. The strength of interactions evolves in time following a simple structural approach. The large-scale network structure is the combination of two processes taking place  at evolutionary time scales: (a) new species are engendered from old ones through speciation and (b) environmental or stochastic factors vary the presence and the strength of species-species interactions. 

Unlike our two previous examples, involving unimodal networks, our model addresses bipartite graphs $G=(A,P,E)$ including now two subsets $A(t)$ and $P(t)$ that correspond to two different sets of species. Here, $t$ needs to 
be understood as an evolutionary time step.  Bipartite networks only have links between species of different kind, that is, when there is a mutualistic relationship. The quantitative impact of animals on plants is weighted through the matrix $\omega_{ij}=\omega(A_i \rightarrow P_j)$, which gives the strength of interaction between  animal species $A_i$ and plant species $P_j$. The quantitive approach leads to the use of spectral radius $\rho(\omega)$, which is a more rigorous assessment of nestedness than previous studies (Staniczenko et al., 2013, Fischer and Lindenmayer, 2002). From this weighted network, we can also limit ourselves to the topological network, where $\omega_{ij}=\omega_{ji} = 1$ when a positive interaction exists and zero otherwise.  

A schematic in Fig \ref{fig:mutual}a illustrates the simple set of neutral rules used to evolve mutualistic networks  (Valverde et al., 2018). During speciation events (denoted by S), a new animal species $A_k$ inherits all the mutualistic interactions from its parent species, $\omega_{kj}=\omega_{ij}$ for all $j=1, ..., |P|$ (see left branch in Fig \ref{fig:mutual}a). A similar speciation event takes place for plant species (see right branch in Fig \ref{fig:mutual}a). This splitting process derives two new species from their common ancestor. In the long term, the lineages diverge due to the random redistribution of link weights between parent and daughter species. Link mutations (denoted by D) satisfy two additional requirements. First, there is a maximum value for the sum of weights attached to any species, i. e. $\sum_j \omega_{ji} \le 1$. And second, links whose weight falls below a minimal threshold are always removed, which can lead to the eventual extinction of whole species, i. e. if $\sum_j \omega_{ji}=0$. This combination of rules represents natural resource constraints, e.g., too many animals feeding on the same plant would rapidly deplete it.    

Speciation-divergence reproduces the observed patterns of network architecture, and without assuming any kind of adaptive mechanism.  To validate our predictions, a data base  (IWDB) of $n=25$ weighted mutualistic webs (Fig \ref{fig:mutual}b is an example of real network) is compared against a large set of simulated bipartite networks randomly sampled from the full parameter space. The predicted degree distribution follows the truncated power-law found in empirical data (see Fig \ref{fig:mutual}c) and analytical 
models.  Not only degree heterogeneity emerges from the neutral model,  but all other measurable network 
attributes, including among them nestedness (see Fig. \ref{fig:mutual}d).  But nestedness has been
shown to be a byproduct of the heterogeneity (Jonhson et al., 2013). Given the lack of a fitness function driving selection, the agreement between theory and data supports the non-adaptive nature of the observed 
architecture (Valverde et al., 2018).  Macroevolutionary patterns shape the global architecture of mutualistic networks, while adaptation and selection work on top of universal features derived from general principles of network growth.  
Recent work further support the importance of ecological spandrels in community assembly (Maynard et al. 2018).


%
\section{Discussion: tinkering and beyond}

%

\begin{quote}
\flushright
\em
"No paradox, no progress" \\
Niles Bohr
\end{quote}

\vspace{0.5 cm}

What forces shape complexity? Since networks are the fabric of complex systems, 
a reasonable way of rephrasing the previous question is: what forces shape network 
complexity? Too often, the observation of some particular pattern is interpreted in terms of 
optimal design or some kind of adaptive trait. This is a traditional 
tendency among biologists while looking for explanations under the absence of solid arguments (same can
be said about many physicists with a limited understanding of biology and evolution). As pointed out 
by a number of authors (Alberch 1998, Gould 2002) the fundamental constraints imposed by extensive reuse 
pervades the presence of most network-level structures, and non-adaptive processes 
might dominate many evolutionary processes (Lynch 2007). Tinkering and reuse are examples of those 
non-adaptive mechanisms.

In this paper we have reviewed three relevant examples illustrating this 
conceptual framework that show how duplication and rewiring inevitably occur either as evolutionary 
tinkering or as a consequence of a complexity design horizon. The later provides an unexpected twist to Jacob's 
proposal: the engineer can sometimes act as a tinkerer. This is somewhat a paradoxical situation: 
as a software project keeps growing, complexity also grows in a nonlinear fashion, 
and individual understanding of the system decreases. In parallel, robust patterns also emerge. 
Because of the inevitable tendency 
towards heterogeneous distributions of links resulting from the amplification process, 
several desirable properties emerge {\em for free}: the efficient connectivity associated to the 
presence of hubs, along with the sparse (and thus low-cost in terms of connections) guarantees 
efficient propagation of information among subparts. The modular or nested organisation facilitate 
the development of specialized (but integrated) subparts. Here too division of labour can be 
selected on top of the incipient modular structure. The tinkering scenario thus enhances the 
chances for natural selection to exploit an advantageous network topology. 

The architecture of these complex networks and our knowledge of the microscopic rules 
provide the basis for identifying the origins of web complexity. 
What about other systems, where tinkering might be less relevant? 
Optimality can certainly work in some cases, particularly in relation with 
tree-like structures, global optima can be reached through evolutionary search (West et al.1997) 
or energy minimization principles (Rodriguez-Iturbe and Rinaldo, 2001). 
Most evolutionary paths towards optimality in networks, particularly spatial ones, 
require considering a conflict between efficient communication and wiring costs (Ferrer-i-Cancho and Sol\'e 2003; 
Colizza et al 2004, see Seoane and Sol\'e 2015 and references cited) or conflicts between 
different subparts (Rasskin-Gutman and Esteve-Altava  2014; Esteve-Altava et al. 2015; 
Bruner et al. 2019). In most cases, percolation and 
geometry constraints (along with graph heterogeneity) can play a crucial role in limiting the space of possible networks that 
can exist (Corominas-Murtra et al, 2013; Avena-Koenigsberger et al 2015). 

Optimization and emergence seem to belong to totally different domains, but they are not when 
dealing with evolved biological nets and their technological counterparts (Sol\'e et al 20013). 
This is particularly important when a conflict between two (or more) traits need to be simultaneously 
optimised. This can led to phase transitions where the conflict is resolved along with the emergence of some inevitable network 
organisation features (Sol\'e and Seoane, 2015). An example is provided by an information theory model of emergence of language (Ferrer-i-Cancho and Sol\'e 2003, Seoane and Sol\'e 2018). Language has been a great challenge to scholars from very different disciplines 
and remains one of the most difficult to explain as a major evolutionary transition (szathm\'ary and Smith 1995). In this context, 
as pointed by Tecumseh Fitch, language 'can be seen as a "bag of tricks" pieced together via a process of evolutionary 
tinkering' (Fitch 2010). As it also occurs with some features of the neural cortex, cognitive complexity can emerge out of reuse. 
Does this process create common patterns here too? What kind of universals might be there? Interestingly, here 
too a set of components (words) interacts in complex ways and some striking statistical patterns seem to be universal. This is 
the case of so called Zipf's law. Imagine we count all words that appear within a text and rank them from the most frequent to the least 
one. Zipf's law Is a scaling law, common to all known languages and says that the frequency of a word within a given 
(long) text decays with the inverse of its rank. How is this connected with the way words interact? 

\begin{figure}[t]
\begin{center}
\includegraphics[width=0.45 \textwidth]{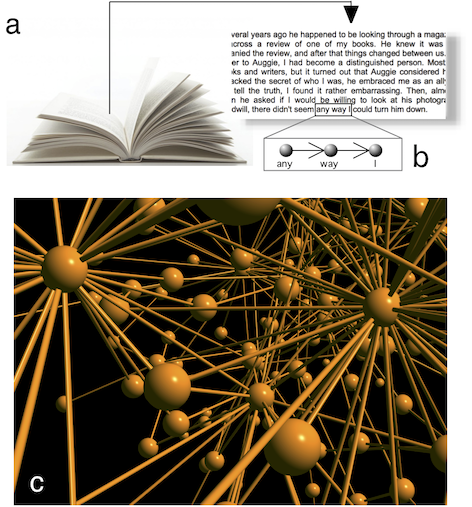}
\caption{Language networks exhibit scale-free structure and several kinds of correlations. 
A simple way of generating a web of words is the use of co-occurrence links. Here two 
words from a given text (a) are connected if they appear one after the other (at least once) within a given sentence (b). 
In (c) we display a zoom to the co-occurrence (purely topological) network obtained from the first 
paragraphs of {\em Augie Wren's Xmas Tale} by Paul Auster.  The resulting web has some marked 
features, such as the lack of connections between hubs (disassortativity) and very short path lengths. 
}
\label{fig:auggie}
\end{center}
\end{figure}

One of the most successful examples of application of network science is the study of 
language as a web of connected words. Specifically, using diverse ways of relating words by their appearance within 
sentences (Ferrer-i-Cancho et al., 2004), semantic relations (Steyvers and Tenenbaum 2005; Motter et al 2002; Sigman and Cecchi, 2002) 
or syntax (Ferrer-i-Cancho and Sol\'e 2004; Sol\'e 2005) it was early shown that the resulting graphs shed 
light on the the functional and evolutionary aspects of their global organisation 
(Sol\'e et al 2010). A very simple illustration is provided by co-occurrence 
graphs obtained from written texts (fig \ref{fig:auggie}a) from which two words are considered 
to be linked if they appear one after the other within (at least) one sentence 
(fig \ref{fig:auggie}b). By using this definition, a scale-free network is obtained (figure 5c). 
Specifically, the frequency distribution of words having $k$ co-occurrences with others was shown to scale as 
$P(k) \sim k^{-\gamma}$. Several interesting properties can be detected, among others 
very short path lengths, high clustering and the lack of links between prepositions and other words acting as 
hubs.

It can be shown that some of these properties emerge for free from a simple model based on 
a conflict between cognitive efforts (Ferrer-i-Cancho and Sol\'e 2003). In a nutshell, language complexity 
would be the result of simultaneously reduce the cost of both speaker and hearer (Zipf 1949) by minimising 
the error on both sides of the communication channel. Using two well-defined quantities from information theory 
as measures of efforts, it was shown that a phase transition takes place when both efforts are equal. At that point, 
Zipf's law emerges along with the underlying network of word-word interactions (Ferrer-i-Cancho et al 2005, Sol\'e 2005; 
Corominas-Murtra et al, 2009, 2011; Salge et al. 2015). 

Tinkering processes are one major driving force in the historical and evolutionary dynamics of complex networks. 
If duplication-divergence rules are responsible for network growth, they deeply constrain the structural outcome, 
creating graph correlations that are measurable and largely universal. As a consequence, order comes for free: 
observable regularities are the expected result of amplification processes.  Two important lessons are of 
relevance to evolutionary dynamics. The first, that convergent patterns found at the network level can be largely 
explained by the mathematics of network growth under reuse of parts. Secondly, a considerable amount of topological 
precursors required to develop adaptive traits and robustness 
are in place with no active selection. These ideas can be extended 
to technological systems, where network organisation has been shown to be irreducible to the individual-level 
decisions based on rational engineering practices (Sol\'e et al. 2013; McNerney et al 2011). This irreducible complexity 
is at the core of the structuralist view of evolution (Alberch 1998; Gould 2002) that has been particularly relevant within the 
context of development. Structuralism holds the view that self-organisation and constraints play a leading role 
in defining the space of the possible, sometimes as a consequence of physical constraints 
(Forgacs and Newman 2005; Oll\'e et al 2016). 
Our view is that topological constraints play a similar role in the context of network complexity.

\begin{acknowledgments}
We would like to thank the members of the Complex Systems Lab for very stimulating and useful discussions. RS thanks Stuart Kauffman and the late Brian Goodwin for many conversations on evolution, complexity and networks. SV thanks Javier Salvador-Calvo, Francisco Jos\'e Garc\'ia-Ojalvo and Jose Paredes for many discussions around the implications of software networks.  Both authors acknowledge the inspiration of John Lokitis. This work has been supported by the Bot\'in Foundation by Banco Santander through its Santander Universities Global Division, a MINECO FIS2015-67616 fellowship and the Santa Fe Institute (RS) and by the Spanish Ministry of Economy and Competitiveness, grant FIS2016-77447-R MINECO/AEI/ FEDER and European Union (to S.V.). This work has also counted with the support of Secretaria d'Universitats i Recerca del Departament d'Economia i Coneixement de la Generalitat de Catalunya and  by the "Mar\'ia de Maeztu" Programme for Units of Excellence in R\&D (MDM- 2014-0445), and from the CERCA Programme of the Generalitat de Catalunya (RS and SV) . 
\end{acknowledgments}

\vspace{0.75 cm}
{\bf \large References}
\vspace{0.25 cm}

\begin{enumerate}

\item
Alberch, P., 1989. The logic of monsters: evidence for internal constraint in development and evolution. Geobios, 22, 21-57.

\item
Albert, R. and Barab\'asi, A.L., 2002. Statistical mechanics of complex networks. Reviews of modern physics, 74, 47.	

\item
Alexander-Bloch, A. F., Gogtay, N., Meunier, D., Birn, R., Clasen, L., Lalonde, F., et al. 2010. 
Disrupted modularity and local connectivity of brain functional networks in childhood-onset schizophrenia. Front. Syst. Neurosci. 4-147. 

\item
Amaral, L.A.N., Scala, A., Barthelemy, M. and Stanley, H.E. 2000. Classes of behavior of small-world networks. Proc. Nat. Acad. Sci. USA, 97, 11149-11152.

\item
Anderson, P.W., 1972. More is different. Science, 177, 393-396.

\item
Arabidopsis Interactome Mapping Consortium, 2011. Evidence for network evolution in an Arabidopsis interactome map. Science, 333, 601-607.

\item
Avena-Koenigsberger, A., Go\~ni, J., Sol\'e, R. and Sporns, O., 2015. Network morphospace. Journal of the Royal Society Interface, 12(103), 20140881.

\item
Banzhaf, W. and Kuo, P.D.  2004. Network motifs in natural and artificial transcriptional regulatory networks. J. Biol. Phys. Chem. 4, 85-92.

\item
Barab\'asi, A.L., 2016. {\em Network science}. Cambridge university press.

\item
Barrat, A., Barth\'elemy, M., Pastor-Satorras, R., and Vespignani, A. 2004. The architecture of complex weighted networks. Proc. Nat. Acad. Sci. USA101(11), 3747-3752.

\item
Bascompte, J., Jordano, P., Meli\'an, C.J. and Olesen, J.M. 2003. The nested assembly of plant-animal mutualistic networks. Proceedings of the National Academy of Sciences USA, 100, 9383-9387.

\item
Bascompte, J. et al. 2006. Asymmetric coevolutionary networks facilitate biodiversity maintenance. Science 312: 431-433.

\item
Bascompte, J., and Jordano, P. 2007. Plant-Animal Mutualistic Networks: The Architecture of Biodiversity. Annu. Rev. Ecol. Evol. Syst. 38, 567-593.

\item
Bashan, A., Bartsch, R. P., Kantelhardt, J. W., Havlin, S., and Ivanov, P. Ch. 2012. Network physiology reveals relations between network topology and physiological function. Nat. Commun. 3-702. 

\item
Bassett, D.S. and Sporns, O., 2017. Network neuroscience. Nature Neuroscience, 20(3), 353.

\item
Bastolla, U., Fortuna, M.A., Pascual-Garc\'ia, A., Ferrera, A., Luque, B., and Bascompte, J. 2009. The architecture of mutualistic networks minimizes competition and increases biodiversity. Nature, 458, 1018-1020.

\item
Boccaletti, S., Latora, V., Moreno, Y., Chavez, M. and Hwang, D.U., 2006. Complex networks: Structure and dynamics. Physics reports, 424,175-308.

\item
Bruner, E., Esteve-Altava, B. and Rasskin-Gutman, D., 2019. 
A network approach to brain form, cortical topology and human evolution. Brain Structure and Function, pp.1-15.

\item
Calabretta R, Nolfi S, Parisi D, Wagner GP. Duplication of modules facilitates the evolution of functional specialization. Artif Life. 2000;6:69?84.

\item
Camacho, J., Guimerà, R. and Amaral, L.A.N., 2002a. Robust patterns in food web structure. Physical Review Letters, 88(22), 228102.

\item
Camacho, J., Guimera, R. and Amaral, L.A.N., 2002b. Analytical solution of a model for complex food webs. Physical Review E, 65(3), 030901.

\item
David, A. S. 1994. Dysmodularity: a neurocognitive model for schizophrenia. Schizophr. Bull. 20, 249-255. 

\item
Caldarelli, G., Higgs, P. G. and McKane, A. J. 1998. Modelling Coevolution in Multispecies Communities. J. Theor. Biol. 193, 345-358.

\item
Christensen, K., Di Collobiano, S. A., Hall, M., and Jensen, H. J. 2002. Tangled nature: a model of evolutionary ecology. J. Theor. Biol. 216, 73-84.

\item
Clune, J., Mouret, J. B. and Lipson, H. 2013. The evolutionary origins of modularity. Proc. R. Soc. B 280, 20122863.

\item
Colizza, V., Banavar, J.R., Maritan, A. and Rinaldo, A., 2004. Network structures from selection principles. Physical review letters, 92(19), p.198701.

\item
Colizza, V., Flammini, A., Maritan, A. and Vespignani, A., 2005. 
Characterization and modeling of protein-protein interaction networks. Physica A 352, 1-27.

\item
Concas, G., Marchesi, M., Pinna, S., Serra, N. 2007. Power-Laws in a Large Object-Oriented Software System. IEEE Trans. Soft. Eng. 33 (10), 687-708.

\item
Concas, G., Marchesi, M., Monni, C., Orr\`u, M.  and Tonelli, R. 2017. Software Quality and Community Structure in Java Software Networks.  International Journal of Software Engineering and Knowledge Engineering 27 (7), 1063-1096. 

\item
Corominas-Murtra, B., Valverde, S. and Sol\'e, R., 2009. 
The ontogeny of scale-free syntax networks: phase transitions in early language acquisition. 
Advances in Complex Systems, 12(03), 371-392.

\item
Corominas-Murtra, B., Fortuny, J. and Sol\'e, R.V., 2011. Emergence of Zipf's law in the evolution of communication. 
Physical Review E, 83(3), 036115.	

\item
Corominas-Murtra, B., Go\~ni, J., Sol\'e, R. and Rodriguez-Caso, C., 2013. 
On the origins of hierarchy in complex networks. Proceedings of the National Academy of Sciences, 110, 13316-13321.

\item
DeFelipe, J., 2006. Brain plasticity and mental processes: Cajal again. Nature Reviews Neuroscience, 7, 811.

\item
Dennett, D. C. 1995. {\em Darwin's dangerous idea}. Simon and Schuster, New York.

\item
Dorogovtsev, S.N. and Mendes, J.F.F. 2002. Evolution of Random Networks. Adv. Phys.  51, 1079-1187.

\item
Dunne, J. A., Williams, R. J., and Mart\'inez, N. D. 2002. Food-web structure and network theory: The role of connectance and size. PNAS 99 (20), 12917-12922.

\item
Esteve-Altava, B., Diogo, R., Smith, C., Boughner, J.C. and Rasskin-Gutman, D., 2015. 
Anatomical networks reveal the musculoskeletal modularity of the human head. Sci. rep. 5, 8298.

\item
Estrada, E., 2012. The structure of complex networks: theory and applications. Oxford University Press.

\item
Ferrer-i-Cancho, R. and Sol\'e, R., 2001. The small world of human language. Proceedings of the Royal Society of London. B, 268(1482), 2261-2265.

\item
Ferrer-i-Cancho, R. and Sol\'e, R.., 2003. Optimization in complex networks. In Statistical mechanics of 
complex networks (pp. 114-126). Springer, Berlin, Heidelberg.

\item
Ferrer-i-Cancho, R., Sol\'e, R. V. and K\"ohler, R., 2004. Patterns in syntactic dependency networks. Physical Review E, 69, 051915.

\item
Ferrer-i-Cancho, R., Riordan, O. and Bollob\'as, B., 2005. The consequences of Zipf's law for syntax and symbolic reference. 
Proceedings of the Royal Society B: Biological Sciences, 272(1562), 561.

\item
Fischer, J. and Lindenmayer, D. B. 2002. Treating the nestedness temperature calculator as a ?black box? can lead to false conclusions. Oikos 99(1), 193-199.

\item
Fitch, W.T., 2010. {\em The evolution of language}. Cambridge University Press.

\item
Forgacs, G. and Newman, S.A., 2005. Biological physics of the developing embryo. Cambridge University Press.

\item
Friedman, D. P., Wand, M., and Haynes, C. T. 2008. {\em Essentials of Programming Languages} 3rd ed. MIT Press. 

\item
Gell-Mann, M., 1995. {\em The Quark and the Jaguar: Adventures in the Simple and the Complex}. Macmillan.

\item
Godwin, D., Barry, R. L., and Marois, R. 2015. Breakdown of the brain?s functional network modularity with awareness. Proc. Natl. Acad. Sci. U.S.A. 112, 3799-3804. 

\item
Goldenfeld, N., 2018. Lectures on phase transitions and the renormalization group. CRC Press.

\item
Goodwin, B., 1994. {How the leopard changed its spots: The evolution of complexity}. Princeton University Press.

\item
Gould, S.J. (2002) {\em The Structure of Evolutionary Theory}, Harvard University Press. 

\item
Gould, S.J. and Lewontin, R.C. (1979) The Spandrels of San Marco and the Panglossian Paradigm: a critique of the Adaptationist Programme. Proc. R. Soc. B 205, 581-598.

\item
Gross, J.L. and Yellen, J., 2005. Graph theory and its applications. CRC press.

\item
Itzkovitz, S., Milo, R., Kashtan, N., Ziv, G., Alon, U. 2003. Subgraphs in random networks. Phys Rev E 68:026127.

\item
Jacob, F. 1977. Evolution and Tinkering. Science 196, 1161-166.

\item
James, A., Pitchford, J. W. and Plank, M. J. 2012. Disentangling nestedness from models of ecological complexity. Nature 487, 227-230.

\item
Jonhson, S., Dominguez-Garcia, V. and Mu\~noz, M. A. 2013. Factors determining nestedness in complex networks. PloS one, 8(9), e74025.

\item
Jordano, P. et al. 2003. Invariant properties in coevolutionary networks of plant-animal interactions. Ecol. Lett. 6: 69-81.

\item
Kauffman, S.A., 1993. {\em The origins of order: Self-organization and selection in evolution}. Oxford U. Press.

\item
Kauffman, S.A. and Levin, S. 1987. Towards a general theory of adaptive walks on rugged landscapes. J Theor Biol 128, 11-45.

\item
Kauffman, S.A. and Johnsen, J. 1991. Coevolution to the edge of chaos: coupled fitness landscapes, poised states and coevolutionary avalanches. J. Theor. Biol. 149, 467-505.

\item
Krapivsky, P.L. and Redner, S., 2005. Network growth by copying. Physical Review E, 71(3), 036118.

\item
Krinke J. 2008. Is cloned code more stable than non-cloned code? Eighth IEEE international working conference on source code analysis and manipulation,  57-66.

\item
Lakos, J., 1996. {\em Large Scale C++ Software Design} Addison-Wesley, New York.

\item
Lipson H, Pollack JB, Suh NP. On the origin of modular varia- tion. Evolution. 2002;56:1549?56.

\item
Loreau, M. 2010. Linking biodiversity and ecosystems: towards a unifying ecological theory. Phil. Trans. R Soc. B 365, 49-60.

\item
Louridas, P., Spinellis, D., and Vlachos, V. 2008. Power Laws in Software. ACM Trans. Soft. Eng. Meth. 18(1), 1-26.

\item
Lynch, M. 2007. The evolution of genetic networks by non-adaptive processes. Nature Rev. Genet. 8, 803-813.

\item
Margalef, R., 1968. Perspectives in ecological theory. University of Chicago Press.

\item
May, R.M., 1973. {\em Stability and complexity in model ecosystems}. Princeton university press.

\item
Maynard, D.S., Serván, C.A. and Allesina, S., 2018.  Network spandrels reflect ecological assembly. Ecology letters, 21(3), 324-334.

\item
Mazurie, A. et al. 2005. An evolutionary and functional assessment of regulatory network motifs. Genome Biol. 6, R35.

\item
Meunier, D., Lambiotte, R., and Bullmore, E. T. 2010. Modular and  hierarchically modular organization of brain networks. Front. Neurosci. 4-200. 

\item
McComb, W. D. 2008. {\em Renormalization Methods: A Guide For Beginners}. Oxford U. Press.

\item
McGuffee, S.R. and Elcock, A.H., 2010. Diffusion, crowding and protein stability in a dynamic molecular model of the bacterial cytoplasm. PLoS computational biology, 6(3), p.e1000694.

\item
McNerney, J., Farmer, J.D., Redner, S. and Trancik, J.E., 2011. Role of design complexity in technology improvement. 
Proceedings of the National Academy of Sciences, 108, 9008-9013.

\item
Memmott, J. et al. 2004. Tolerance of pollination networks to species extinctions. Proc. R. Soc. B 271, 2605-2611.

\item
Middendorf, M., Ziv, E. and Wiggins, C.H., 2005. Inferring network mechanisms: the Drosophila melanogaster protein interaction network. Proceedings of the National Academy of Sciences, 102(9), pp.3192-3197.

\item
Milgram, S., 1967. The small world problem. Psychology today, 2(1), 60-67.

\item
Milo, R., Shen-Orr, S., Itzkovitz, S., Kashtan, N., Chklovskii, D., Alon, U. 2002. Network motifs: simple building blocks of complex networks. Science 298, 824-827.

\item
Montoya, J.M. and Sol\'e, R.V. 2002 Small world patterns in food webs. J. Theor. Biol. 214, 405-412.

\item
Montoya, J. M., Pimm, S. and Sol\'e, R. V. 2006. Ecological networks and their fragility. Nature 442 (7100), 259-264.

\item
Motter, A.E., De Moura, A.P., Lai, Y.C. and Dasgupta, P., 2002. Topology of the conceptual network of language. Physical Review E, 65(6),065102.

\item
Nehaniv, C. L., Hewitt, J., Christianson, B., Wernick, P. 2006. What software evolution and biological evolution don?t have in common.  Second international IEEE workshop on software evolvability. Philadelphia: IEEE Computer Society, 58-65.

\item
Newman, M. E. J. and Palmer, R. 2003. {\em Modelling extinction.} Oxford U. Press. New York.

\item
Newman, M.E., 2003. The structure and function of complex networks. SIAM review, 45(2),167-256.

\item
Olesen, J. et al. 2006. The smallest of all worlds: pollination networks. Ecological networks and their fragility.  J. Theor. Biol. 240, 270-276.

\item
Olle-Vila, A., Duran-Nebreda, S., Conde-Pueyo, N., Montanez, R. and Sole, R., 2016. 
A morphospace for synthetic organs and organoids: the possible and the actual. Integrative Biology, 8(4), 485-503.

\item
Potanin, A., Noble, J.,  Frean, M.,  and Biddle, R. 2005. Scale-free geometry in OO programs. Commun. ACM, 48(5), 99-103.

\item
Pressman, R. S. 2009. {\em Software engineering: a practitioner's approach}, 7th ed. Boston, MA: McGraw-Hill.

\item
Raff RA. The shape of life. Chicago: Chicago U. Press; 1996.

\item
Rasskin-Gutman, D. and Esteve-Altava, B., 2014. 
Connecting the dots: anatomical network analysis in morphological EvoDevo. 
Biological Theory, 9(2), 178-193.

\item
Rodriguez-Caso C., Medina M. A., Sol\'e, R.V. 2005. Topology, tinkering and evolution of the human transcription factor network. FEBS J. 272 (24), 6423-6434.

\item
Rodriguez-Iturbe, I. and Rinaldo, A., 2001. Fractal river basins: chance and self-organization. Cambridge University Press.

\item
Roth, C., Rastogi, S., Arvestad, L., Dittmar, K., Light, S., Ekman, D. and Liberles, D.A., 2007. Evolution after gene duplication: models, mechanisms, sequences, systems, and organisms. Journal of Experimental Zoology Part B: Molecular and Developmental Evolution, 308(1), pp.58-73.

\item
Saavedra, S., Rohr, R. P., Olesen, J. M., and Bascompte, J. 2016. Nested species interactions promote feasibility over stability during the assembly of a pollinator community. Ecology and evolution. 6, 997-1007.

\item
Seoane, L.F. and Sol\'e, R., 2015. Phase transitions in Pareto optimal complex networks. Physical Review E, 92, 032807.

\item
Sigman, M. and Cecchi, G.A., 2002. Global organization of the Wordnet lexicon. Proceedings of the National Academy of Sciences, 99(3),1742-1747.

\item
Sol\'e, R., 2005. Language: Syntax for free?. Nature, 434, .289-290.

\item
Sol\'e, R. and Goodwin, B. C., 2001. {\em Signs of life: how complexity pervades biology}. Basic Books. 

\item
Sol\'e, R.V., Corominas-Murtra, B., Valverde, S. and Steels, L., 2010. Language networks: 
Their structure, function, and evolution. Complexity, 15(6), 20-26.

\item
Sol\'e, R.V. and Manrubia, S. C. 1995. Extinctions and self-organised criticality in a model of large-scale evolution. Phys. Rev E 54, R42-R46.

\item
Sol\'e, R. V., and Montoya, J. M. 2001. Complexity and fragility in ecological networks.  Proc. Roy. Soc. Lond. Ser B 268, 2039-2045.

\item
Sol\'e, R.V., Pastor-Satorras, R., Smith, E., Kepler, T.  2002. A model of large-scale proteome evolution. Adv. Complex Syst. 5, 43-54.

\item
Sol\'e, R. V., Ferrer-i-Cancho, R., Montoya, J. and Valverde, S. 2003. Selection, Tinkering, and Emergence in Complex Networks. Complexity 8(1), 20-33.

\item
Sol\'e R. V., and Valverde, S. 2006. Are network motifs the spandrels of cellular complexity? Trends Ecol. Evol. 21(8), 419-422.

\item
Sol\'e, R. V., and Bascompte, J. 2006. {\em Self-organization in complex ecosystems}. Princeton U. Press. 

\item
Sol\'e, R. V., Valverde, S., Casals, M.R., Kauffman, S.A., Farmer, D. and Eldredge, N., 2013. 
The evolutionary ecology of technological innovations. Complexity, 18(4),15-27.

\item
Sol\'e, R.V. and Seoane, L.F., 2015. Ambiguity in language networks. The Linguistic Review, 32, 5-35.

\item
Sol\'e, R.V., and Valverde, S. 2008. Spontaneous emergence of modularity in cellular networks. Journal of the Royal Society Interface 5 (18), 129-133. 

\item
Sporns, O., Chialvo, D.R., Kaiser, M. and Hilgetag, C.C., 2004. Organization, development and function of complex brain networks. Trends in cognitive sciences, 8(9), 418-425.

\item
Staniczenko, P. P. A., Kopp, J. C., and Allesina, S. 2013. The ghost of nestedness in ecological networks. Nature Comm. 4, 1391.  

\item
Stanley, H.E., 1971. {\em Phase transitions and critical phenomena}. Clarendon Press, Oxford.
Vancouver	

\item
Stauffer, D. and Aharony, A., 2014. {\em Introduction to percolation theory}. Taylor and Francis.

\item
Stein, D.L. and Newman, C.M., 2013. {\em Spin glasses and complexity}. Princeton University Press.
Vancouver	

\item
Steyvers, M. and Tenenbaum, J.B., 2005. The large-scale structure of semantic networks: 
Statistical analyses and a model of semantic growth. Cognitive science, 29(1), 41-78.

\item
Stone, L., Simberloff, D. and Artzy-Randrup, Y., 2019. Network motifs and their origins. PLoS computational biology, 15, e1006749.

\item
Subelj, L. and Bajec, M. 2011. Community structure of complex software systems: Analysis and applications.
Phys. A: Stat. Mech. Appl. 390 (16), 2968-2975.

\item
Szathm\'ary, E. and Smith, J.M., 1995. The major evolutionary transitions. Nature, 374, 227.

\item
Teichmann, S.A. and Babu, M.M., 2004. Gene regulatory network growth by duplication. Nature genetics, 36, 492.

\item
V\'azquez, A., Flammini, A., Maritan, A. and Vespignani, A., 2003. Modeling of protein interaction networks.  Complexus, 1, 38-44.

\item
Valverde, S., Ferrer-i-Cancho, R., and Sol\'e, R. V. 2002. Scale-free Networks from Optimal Design. Europhys. Lett. 60, 512-517.

\item
Valverde, S., and Sol\'e, R. V. 2007. Hierarchical small worlds in software architecture. Dynam. Cont. Discr. Impul. Syst. Ser. B 14, 1-11. 

\item
Valverde, S., and Sol\'e, R. V. 2005. Network motifs in computational graphs: A case study in software architecture. Phys. Rev. E 72, 026107.

\item
Valverde, S., and Sol\'e, R. V. 2005. Logarithmic growth dynamics in software networks. Europhys. Lett. 72, 858-864. 

\item
Valverde, S., 2007. Crossover from endogenous to exogenous activity in open-source software development. Europhys. Lett. 77, 20002.

\item
Valverde, S., and Sol\'e, R. V. 2009. Motifs in graphs.  Encyclopedia of Complexity and Systems Science (Robert A. Meyers, ed.), Springer Science+Business Media, New York, 5692-5702. 

\item
Valverde, S., and Sol\'e, R. V. 2015. Punctuated equilibrium in the large-scale evolution of programming languages. J. R. Soc. Interface 12, 20150249.

\item
Valverde, S. 2016. Major transitions in information technology. Phil. Trans. R. Soc. B 371, 20150450.

\item
Valverde, S. 2017. Breakdown of Modularity in Complex Networks. Front. Physiol. 8, 497. 

\item
Valverde, S., Pi\~nero, J., Corominas-Murtra, B., Montoya, J., Joppa, L. and Sol\'e, R. V. 2018. The architecture of mutualistic networks as an evolutionary spandrel. Nat. Ecol. Evol. 2, 94-99. 

\item
Vazquez, A. 2003. Growing network with local rules: Preferential attachment, clustering hierarchy, and degree correlations. Phys. Rev E67, 056104.

\item
Vázquez, A., Flammini, A., Maritan, A. and Vespignani, A., 2003. Modeling of protein interaction networks. Complexus, 1, 38-44.

\item
Wagner GP, Pavlicev M, Cheverud JM. The road to modularity. Nature Rev Genet. 2007;8:921?31.

\item
Watts, D. J. and Strogatz, S. H. 1998. Collective dynamics of "small-world" networks. Nature 393: 440-442.

\item
Widder, S., Solé, R. and Macía, J., 2012. Evolvability of feed-forward loop architecture biases its 
abundance in transcription networks. BMC systems biology, 6(1), 7.

\item
Williams, R. J., and Martinez, N. D. 2000. Simple rules yield complex food webs. Nature 404, 180-183.

\item
Yu, I., Mori, T., Ando, T., Harada, R., Jung, J., Sugita, Y. and Feig, M., 2016..
 Biomolecular interactions modulate macromolecular structure and dynamics in atomistic model of a bacterial cytoplasm. Elife 5, e19274.
 
\item
 Zipf, G.K., 1949. Human behavior and the principle of least effort. Addison-Wesley. 

\end{enumerate}

\end{document}